\documentclass[onecolumn,prl,aps,showpacs,superscriptaddress,floatfix,nofootinbib,10pt]{revtex4}

\usepackage{amsfonts,amsmath,amssymb}
\usepackage{graphicx}
\usepackage{color}
\usepackage[normalem]{ulem}
\usepackage{multirow}
\usepackage{amsfonts}
\usepackage{afterpage}
\usepackage[toc,page]{appendix}
\usepackage[ocgcolorlinks,colorlinks=true,linkcolor=blue,citecolor=red]{hyperref}
\usepackage{latexsym}
\usepackage{amsfonts}
\usepackage{algpseudocode}
\usepackage{amsthm}
\usepackage{mathrsfs}
\usepackage{natbib}
\usepackage{color,verbatim}
\usepackage{psfrag}

\DeclareMathOperator{\Tr}{tr}

\newcommand{\be}{\begin{eqnarray}}
\newcommand{\ee}{\end{eqnarray}}
\DeclareMathAlphabet{\mathrsfs}{U}{rsfs}{m}{n}
\DeclareMathAlphabet{\mathpzc}{OT1}{pzc}{m}{it}
\DeclareMathAlphabet{\matheus}{U}{eus}{m}{n}
\DeclareMathAlphabet{\mathbbold}{U}{bbold}{m}{n}

\newcommand{\ket}[1]{|#1\rangle}
\newcommand{\bra}[1]{\langle#1|}

\begin{document}
\title{Communication games reveal preparation contextuality}

\author{Alley Hameedi}\thanks{A. H. and A. T. contributed equally for this project.}
\affiliation{Department of Physics, Stockholm University, S-10691 Stockholm, Sweden}

\author{Armin Tavakoli}\thanks{A. H. and A. T. contributed equally for this project.}
\affiliation{Department of Physics, Stockholm University, S-10691 Stockholm, Sweden}
\affiliation{Groupe de Physique Appliqu\'ee, Universit\'e de Gen\`eve, CH-1211 Gen\`eve, Switzerland}

\author{Breno Marques}
\email{bmgt@if.usp.br}
\affiliation{Instituto de F\'isica, Universidade de S\~ao Paulo, 05315-970 S\~ao Paulo, Brazil}

\author{Mohamed Bourennane}
\affiliation{Department of Physics, Stockholm University, S-10691 Stockholm, Sweden}


\date{\today}


\begin{abstract}
A communication game consists of distributed parties attempting to jointly complete a task with restricted communication. Such games are useful tools for studying limitations of physical theories. A theory exhibits preparation contextuality whenever its predictions cannot be explained by a preparation noncontextual model. Here, we show that communication games performed in operational theories reveal the preparation contextuality of that theory. For statistics obtained in a particular family of communication games, we show a direct correspondance with correlations in space-like separated events obeying the no-signaling principle. Using this, we prove that all mixed quantum states of any finite dimension are preparation contextual. We report on an experimental realization of a communication game involving three-level quantum systems from which we observe a strong violation of the constraints of preparation noncontextuality.
\end{abstract}


\pacs{03.67.Hk,
03.67.-a,
03.67.Dd}

\maketitle

\textit{Introduction.---} Communication games are tools by which one can study fundamental limiting features of physical theories in terms of their ability to process information \cite{IC09, GH14, BB15}. In these games, a number of parties intend to jointly solve a task despite the amount and type of communication being constrained by some rules.  Thus, the task can only be solved with some probability, which depends on the theory by which they are assumed to operate. Therefore, communication games are frequent tools for identifying and quantifying quantum advantages over classical theories \cite{GB10, HG12, AB12, PB11, LP12, AB14, T15}.

Interestingly, there are known examples of communication games in which the better-than-classical performance constitutes a certificate of the system lacking a preparation noncontextual ontological model \cite{SB09, CK16, A16}. An ontological model is a way of explaining the physics of an operational theory, by assuming that there are independent and objective (ontic) states subject to experiment. However, specifying a preparation does not necessarily specify the ontic state. A preparation may be represented by a distribution $\mu$ over the ontic states. Let two preparations, $P_1$ and $P_2$, associated to distributions $\mu_1$ and $\mu_2$ be indistinguishable, i.e., satisfy $p(b|P_1,M)=p(b|P_2,M)$ for any measurement $M$ with outcome $b$. The assumption of preparation noncontextuality asserts no additional features (called \textit{contexts}) influence the physics of the preparations, and therefore asserts that both preparations have equivalent representation in the ontological model; $\mu_1=\mu_2$ \cite{Sp05}. If a theory does not satisfy this assumption, it is said to be preparation contextual. Preparation contextuality has been shown relevant for many foundational topics \cite{BB15, S08, P14, LM13, KS15}.

Here, we show that the performance of an operational theory in communication games constitutes a certificate of that theory exhibiting preparation contextuality. Specifically, we introduce communication constraints which keep the receiver oblivious about subsets of the information held by the sender. Preparation noncontextuality imposes a bound on the performance of any communication game executed under such an obliviousness constraint. This bound is violated by preparation contextual theories. Subsequently, we show how to understand no-signaling correlations from space-like separated measurements (perhaps violating a Bell inequality) through a subclass of communication games. In particular, we find that quantum preparation contextuality manifested in communication games imposes a quantitative bound on quantum nonlocality (i.e., Bell inequality violations). Furthermore, we apply this result to resolve an open problem in this field; which quantum states are preparation contextual? We show that all mixed quantum states in any finite dimension are preparation contextual. Finally, we present an experimental implementation of a quantum strategy in a specific communication game, inpired by the Collins-Gisin-Linden-Massar-Popescu (CGLMP) Bell inequality, in which three-level quantum systems are communicated. We certify a large violation of a preparation noncontextuality inequality.

\textit{Communication games.---}
In a two-player communication game, a party Alice (Bob) holds a set of data denoted $x\in I_A$ ($y\in I_B$) with associated probability distribution $p_{A}(x)$ ($p_B(y)$). Alice encodes $x$ by preparing a state which is sent to Bob, who attempts to decode it with a measurement labeled $y$. This returns an outcome $b$. Subsequently, a payoff $\mathcal{C}_{x,y}^b\in \mathbb{R}$ is awarded. The average payoff earned by the partnership is
\begin{equation}\label{payoff}
	A\equiv \sum_{x\in I_A}\sum_{y\in I_B}\mathcal{C}_{x,y}^b p_{A}(x)p_B(y)p(b|x,y).
\end{equation} 
Eq.\eqref{payoff} quantifies the \textit{performance} in the game.  However, the content of Alice's communication to Bob is restricted by some communication constraints. These ensure that the game is nontrivial, i.e., Alice cannot simply send $x$ to Bob. A suitable choice of these constraints enables the connection to tests of preparation contextuality.

\textit{Communication games as tests of preparation contextuality.---} 
An operational theory is said to be preparation noncontextual \cite{Sp05} if operationally equivalent preparations imply equivalent distributions over the ontic states:
\begin{equation}\label{noncontex}
	\forall y~\forall b: p(b|x,y)=p(b|x',y)\Rightarrow p(\lambda|x)=p(\lambda|x'),
\end{equation}
where $\lambda$ is a hidden variable, $x$ and $x'$ are two preparations and $y$ denotes a measurement. 

We will now define a class of communication constraints which enables a connection to the premise of Eq.\eqref{noncontex}. The assumption of preparation noncontextuality then leads to a preparation noncontextuality inequality in which the performance in the communication game is the operator.

Construct $L$ subsets of the space $I_A$; $S_k\subset I_A$  for $k=1,\ldots, L$. Now, choose communication constraints as follows: impose an \textit{obliviousness} constraint
\begin{equation}\label{obliv}
\forall y,b,k,k': \frac{1}{q_{k}}\sum_{x\in S_k} p(x|b,y)=\frac{1}{q_{k'}}\sum_{x\in S_{k'}} p(x|b,y).
\end{equation} 
Here $q_{k}=p(x\in S_k)=\sum_{x\in S_k}p_A(x)$ serves as a normalization. In words, Eq.\eqref{obliv} states that no matter the performed measurement and observed outcome; Bob gains no information, as compared to what he knew before communication, about to which set $S_k$  the data $x$ of Alice belongs. Let us now apply Bayes' rule to the above summands: $p(x|b,y)=p(b|x,y)p(x|y)/p(b|y)$. Since $x$ and $y$ are independent, Eq.\eqref{obliv} becomes
\begin{equation}\label{ob}
\forall y~\forall b:\sum_{x\in S_k}p(b|x,y)\frac{p_A(x)}{q_{k}}=\sum_{x\in S_{k'}}p(b|x,y)\frac{p_A(x)}{q_{k'}}.
\end{equation}
Note that each side is a convex combination since $\{p_A(x)/q_{k}\}_{x\in S_k}$ is a probability distribution  over the set $S_k$. Now, note that the probability that the outcome $b$ was obtained from a measurement on a preparation associated to $S_k$ is the convex mixing of its constitutes: $p(b|x\in S_k,y)=
 \sum_{x\in S_k} p(b|x,y)p_A(x)/q_{k}$. Similarly, the distribution of the hidden variable is $p(\lambda|x\in S_k)=\sum_{x\in S_k} p(\lambda|x)p_A(x)/q_{k}$. Putting it all together, we have $\forall y\forall b: p(b|x\in S_k,y)=p(b|x\in S_{k'},y)$, which takes the form of the premise of the preparation noncontextuality statement in Eq. \eqref{noncontex}. Thus, preparation noncontextuality imposes that $p(\lambda|x\in S_k)=p(\lambda|x\in S_{k'})$. Using Bayes' rule we find that $p(x\in S_k|\lambda)/q_{k}=p(x\in S_{k'}|\lambda)/q_{k'}$. This means that despite knowledge of the hidden variable Eq.\eqref{obliv} remains satisfied. 

Given any $\lambda$, Alice encodes $x$ classically knowing that the oblivioussness constraint is satisfied. Therefore, the preparation noncontextual bound $p^{pnc}$ of Eq.\eqref{payoff} is obtained from maximizing Eq.\eqref{payoff} over all classical encodings respecting the obliviousness constraint. Hence, $A \leq p^{pnc}$ is a preparation noncontextuality inequality. $\blacksquare$

Clearly, for a given communication game there is a plethora of ways in which one can choose the obliviousness constraint and construct the associated preparation noncontextuality inequality. In what follows we will examine some interesting cases of the presented framework.

\textit{Communication games based on Bell inequalities.---} 
Consider a general bipartite Bell experiment in which Alice and Bob share a two-particle state with each of them choosing measurements $X \in\{1,\ldots,m_A\}$, for some positive integer $m_A$, and $Y \in\{1,\ldots,m_B\}$, for some positive integer $m_B$, sampled from a distribution $p_A(X)$ and $p_B(Y)$ respectively. Each measurement returns an outcome $a,b\in \{1,\ldots,d\}$. From the resulting  probability distribution $p(a,b|X,Y)$, one constructs the a general Bell inequality 
\begin{equation}\label{be}
	I_b\equiv \sum_{abXY} \mathcal{C}_{X,Y}^{a,b} p_A(X)p_B(Y)p(a,b|X,Y)\leq C,
\end{equation}
where $C$ is the local realist bound, and $\mathcal{C}_{X,Y}^{a,b}$ are real coefficients. 

In the following, we construct a family of communication games and obliviousness constraints inspired by such Bell experiments. Alice is given inputs $(x,x_0)\in \{1,\ldots,m_A\}\times \{1,\ldots,d\}$ admiting the distribution $p(x_0,x)=p_g(x_0|x)p_A(x)$, with $p_A(x=i)=p_A(X=i)$ whereas $p_g(x_0|x)$ is yet to be specified. Bob has an input $y\in \{1,\ldots,m_B\}$ with distribution $p_B(y=i)=p_B(Y=i)$. The inputs $(x_0,x,y)$ in the communication game respectively correspond to $(a,X,Y)$ in the Bell experiment. Having received Alice's communication, Bob earns a payoff $\mathcal{C}_{x,y}^{x_0,b}$ if he outputs $b$ given a measurement of $y$ and that Alice held $(x_0,x)$. The performance is written
\begin{equation}\label{ig}
I_g(\{p_g(x_0|x)\}_x)\equiv \sum_{x_0xyb} \mathcal{C}_{x,y}^{x_0,b} p_A(x)p_B(y)p_g(x_0|x)p(b|x_0,x,y).
\end{equation} 
Notice that for every choice of $\{p_g(x_0|x)\}_x$, we have a different communication game. 

Alice's communication must satisfy the following obliviousness constraint. Partition Alice's $m_Ad$ possible inputs into $m_A$ sets each containing $d$ elements; we define $S_k=\{x_0x|x=k\}$ for $k=1,\ldots, m_A$. The obliviousness constraint requires that Bob gains no information about to which $S_k$ the data $(x_0,x)$ belongs. Inserting this into Eq.\eqref{ob} with $q_k\!=\!p_A(x\!=\!k)$, and using Bayes' rule we obtain
\begin{equation}\label{nosig}
\forall b,y,k,k':\sum_{x_0=1}^{d}p(x_0,b|x=k,y)=\sum_{x_0=1}^{d}p(x_0,b|x=k',y).
\end{equation}
This constraint is an analogy of the directed no-signaling principle imposed by special relativity on correlations in space-like separated measurement events: the probability of Bob's outcome marginalized over Alice's input $x_0$ is independent of Alice's other input $x$. One needs only to relabel $x_0$ by $a$, and $(x,y)$ by $(X,Y)$ to recover the corresponding statement in Bell experiments. 

On the one hand, imagine we run a Bell experiment and achieve some value of $I_b$. Using Bayes' theorem and the obliviousness constraint \eqref{nosig}, it is straightforwardly shown that if we choose the communication game in which $p_g$ coincides with the observed marginals of Alice, $p(a|X)$, one finds $I_g=I_b$. We explicitly consider the case of quantum theory. In a Bell experiment, when Alice performs her measurement $X$ she renders Bob's local state in one of $d$ possible states labeled $\varrho^X_l$ for $l=1,\ldots,d$. The probability of Bob's local state being $\varrho^X_l$ is the probability of Alice obtaining outcome $l$, i.e., $p(a=l|X)$. No-signaling implies that the average state of Bob is independent of the measurement choice $X$ of Alice. We associate for every $X$ the set $\{\varrho^X_l\}_{l=1}^{d}$ to the states in $S_X$ prepared by Alice in our communication game. As shown, these will necessarily satisfy the obliviousness constraint \eqref{nosig} while by construction returning the same performance in the communication game \eqref{ig} as in the Bell experiment, namely $I_g=I_b$.

On the other hand, imagine we have not specified $p_g(x_0|x)$. Let $\lambda$ index all functions $f_\lambda(x):\{1,\ldots,m_A\}\rightarrow\{1,\ldots,d\}$. By choosing a suitable probability distribution $\mu(\lambda)$, we can write $p_g(x_0|x)=\sum_{\lambda}\mu(\lambda)D_A(x_0|x\lambda)$ where $D_A(x_0|x\lambda)=\delta_{f_\lambda(x),x_0}$. Alice then communicates $\lambda$, which contains no information about $x$, to Bob who decodes the message using some strategy $D_B$. We find
\begin{equation}
I_g=\sum_{x_0xyb}\mathcal{C}_{x,y}^{x_0,b}p_A(x)p_B(y)\sum_{\lambda}\mu(\lambda)D_A(x_0|x\lambda)D_B(b|y\lambda).
\end{equation} 
This is precisely the notion of local realist models for the Bell experiment \eqref{be}. Hence, if we choose $p_g(x_0\lvert x)$ such that there there is a local hidden variable strategy that both has, i) $p_g(x_0\lvert x)=p(a\lvert X)$ as a marginal of Alice, and ii) saturates the local realist bound $C$ of \eqref{be}, the preparaption noncontextuality inequality $I_g\leq C$ will be tight. Of particular interest is to choose $p_g(x_0\lvert x)$ such that it coincides with the Alice's marginals in a maximal violation of a Bell inequality given some operational no-signaling theory. Then, we assert that $I_g$ can witness a violation of preparation noncontextuality corresponding to the maximal Bell inequality violation.

Note that only very particular obliviousness constraints and communication games retain the analogy to the no-signaling principle through our construction. In \cite{Supplementary} we present a family of games that is not of the type presented in this section. The corresponding preparation noncontextuality inequalities are many-outcome generalizations of the those based on parity-oblivious multiplexing \cite{SB09}.

\textit{Quantum preparation contextuality limits maximal quantum nonlocality.---} If Alice and Bob share entangled states, all mixed states can be prepared on Bob's side by considering the average of his local state computed over the outcomes of Alice obtained from some measurement. Thus, due to our previous discussion, it follows that the maximal quantum violation of a bipartite Bell inequality is a limitation imposed by the preparation contextuality allowed in quantum theory. This generalizes the result of Ref.\cite{BB15} showing this statement for the Clauser-Horne-Shimony-Holt inequality \cite{CHSH69}. We examplify this generalization by shining light on the numerical quantum violations of the preparation noncontextuality inequalities considered in Ref.\cite{A16}. These inequalities were based on communication games which happen to admit an obliviousness constraint of the form considered in the above section. The corresponding Bell inequalities were in fact studied in Ref.\cite{TM16} in a different context. Comparing the numerics for quantum preparation contextuality \cite{A16} and the quantum nonlocality \cite{TM16} one indeed finds that these agree very accurately.

\textit{All mixed states are preparation contextual.---} The maximally mixed quantum state of dimension $d=2,3,4,5$ is known to be preparation contextual \cite{Sp05, A16}. So is every mixed qubit state \cite{M14}. Our mapping between communication games and Bell inequalities allows us to  straightforwardly show that all mixed quantum states of any dimension $d$ are preparation contextual. For this purpose, consider the CGLMP Bell inequality \cite{CGLMP02} which is a bipartite facet Bell inequality with $d$ outcomes for both observers. For any $d$, this inequality can be violated by all pure bipartite entangled states of dimension $d$ \cite{chen}. Hence, all possible mixed quantum states of dimension $d$ can appear as the average state of Bob after either of Alice's measurements. That average state is just the state of Bob's part of the entangled system. Since quantum strategies in the Bell scenario can be mapped to quantum strategies in a communication game (of the form previously discussed) testing preparation contextuality, it follows that all mixed quantum states of dimension $d$ are preparation contextual.

\textit{A specific communication game.---} Let us focus on the CGLMP Bell inequality with $d=3$ and construct the preparation noncontextuality inequality based on the associated communication game. Following our previous discussion, we let Alice hold $x=x_0x\in\{0,1,2\}\times \{0,1\}$ with $p(x_0,x)=1/6$, and Bob hold $y\in\{0,1\}$ with $p(y)=1/2$. In order to satisfy the obliviousness constraint, Alice's communication $\rho_{x_0x}$ must in quantum theory obey $\sum_{x_0=0}^{2}p(x_0|x=0)\rho_{x_00}=\sum_{x_0=0}^{2}p(x_0|x=1)\rho_{x_01}$. Since the preparation noncontextual bound coincides with the local bound of the CGLMP inequality (which achieves its maximal quantum violation with uniform marginals on Alice), our preparation noncontextuality inequality reads 
\begin{equation} \label{ex2}
A_3\equiv \frac{1}{12}\sum_{x_0xyk}(-1)^k p(b=T_k|x_0,x,y) \leq 1/2.
\end{equation}
where $T_k=x_0-(-1)^{x+y+k}k-xy\mod{3}$ for $k=0,1$. The maximal quantum violation of the CGLMP Bell inequality is $A_3=(3+\sqrt{33})/12\approx 0.7287$ \cite{AD02}, which immediately translates into an equal quantum violation of the inequality \eqref{ex2}. In \cite{Supplementary}, we give the details of the corresponding quantum strategy in the communication game.

\textit{Experiment.---} We experimentally confirm the above prediction of quantum preparation contextuality. The experimental implementation of the communication game uses 3-path encoding for preparing qutrits. Single photons are initially prepared in $|H\rangle$ polarization state by the use of polarization fibre controllers in a SMF. The qutrit state is prepared using the two spatial modes of three polarization beam splitters (PBS)(see Fig.\ref{Fig. 1}). The states required for the game, $|\psi_{in}\rangle = \cos(2\chi_1)|0\rangle + \sin(2\chi_1)\sin(2\chi_2)|1\rangle + \sin(2\chi_1)\cos(2\chi_2)|2\rangle$, are prepared by suitably orienting the half-wave plates (HWPs) $\chi_1$ and $\chi_2$. Details are given in \cite{Supplementary}. 

We use a heralded single photon source generating twin photons at 780nm by spontaneous parametric down conversion (SPDC). In this process, a nonlinear crystal type II (BBO) is pumped using a high power femtosecond laser such that a pump photon probabilistically converts into two lower energy photons, called signal and idler. The twin photons pass through a 3 nm filter and are coupled into single mode fibers to have well-defined spatial and spectral properties. A detection of the idler then heralds the signal photon.

The corresponding experimental setup consists of three subsequent interferometers comprising of single photon interferometers between all three paths followed by a stable and compact  Sagnac interferometer, such that while performing a measurement in a given measurement basis, the state is projected into basis vectors of the chosen basis. The protocol requires measurements in the computational basis, and a second basis defined in \cite{Supplementary}. Moreover, state tomography is performed using measurements in four mutually unbiased bases (MUB), so that the total set of measurements is informationally complete \cite{VQT2}. For this purpose, the choice of a given measurement basis is enabled by suitable orientations of the HWPs $\theta_1$, $\theta_2$ and $\theta_3$ (see tables I in \cite{Supplementary}) and by the introduction of a phase ($\phi_i; i \in {1,2,3}$) between the special modes by employing a set of three wave plates QWP-HWP-QWP (phase shifter box) geometries at different tilding positions \cite{hw}.

\begin{figure}[t]
	\begin{center}
		\includegraphics[width=0.50\linewidth]{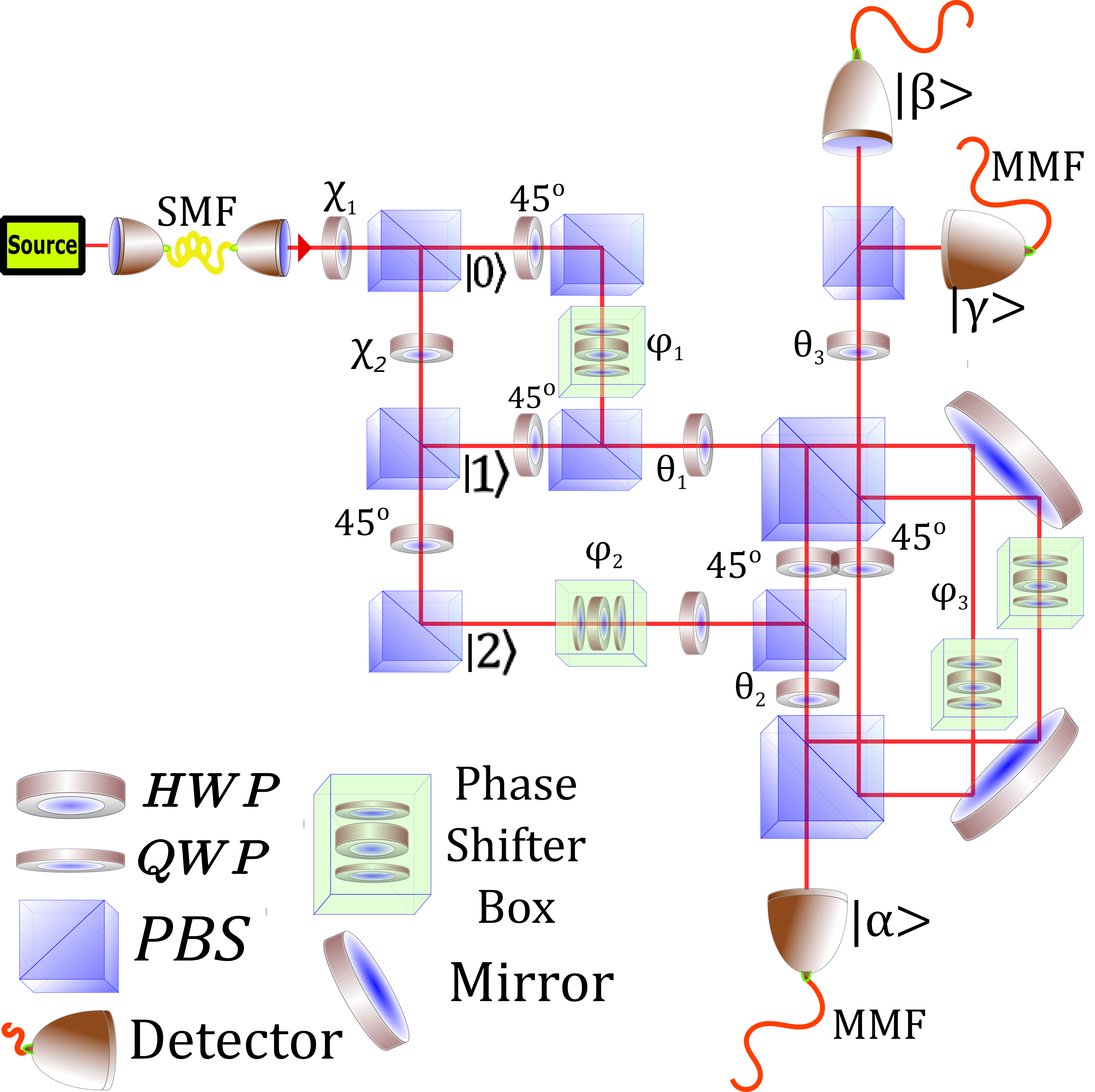}
		\caption{Experimental Setup. Suitable settings of $\chi_1$ and $\chi_2$ allow to produce the desired qutrit states for the task. Measurement basis selection is implemented by appropriate settings of HWPs $\theta_1$, $\theta_2$ and $\theta_3$ and by setting the total experimental phase ($\phi_i; i \in {1,2,3}$) between path modes by employing a phase shifter box (QWP-HWP-QWP) inside the setup. Detection events in detectors $|\alpha\rangle$, $|\beta\rangle$ and $|\gamma\rangle$ are used to obtain the respective probabilities.} 
		\label{Fig. 1}
	\end{center}
\end{figure}

A measurement projects the state onto the basis vectors. These are represented by the spatial modes of the two PBSs in the Sagnac interferometer (denoted by $|\alpha\rangle$, $|\beta\rangle$ and $|\gamma\rangle$). In our experiment, the photons arriving at $|\alpha\rangle$, $|\beta\rangle$ and $|\gamma\rangle$ are collected by multimode fibers that are in turn coupled to single photon silicon avalanche photodiodes (APDs) from Excelitas Technologies with an effective detection efficiency $\eta_d = 0.55$. A home built FPGA based timing system records the coincidence events between the arriving and trigger (idler) photons with a detection time window of $1.7$~ns. The number of detection events at each detector are used to compute the respective probabilities. In each measurement round, approximately $60,000$ photons were detected per second. The measurement time was $10$~s.

From the measured probabilities, we computed $A_3^{pri} \approx 0.7172 \pm 0.0365$ which is in good agreement with the theoretical prediction. We reconstructed the states using variational quantum tomography \cite{VQT1,VQT2} and the experimental results from four MUBs. We found the following fidelities for the six states: $|\psi_{11}\rangle \sim 0.9826$, $|\psi_{12}\rangle \sim 0.9804$, $|\psi_{13}\rangle \sim 0.9893$, $|\psi_{21}\rangle \sim 0.9838$, $|\psi_{22}\rangle \sim 0.9876$, and $|\psi_{23}\rangle \sim 0.9840$. These small imperfections cause the obliviousness constraint not to be perfectly satisfied. Next, we shall see how to overcome this issue.

\textit{Data analysis.---} Ref.\cite{MP16} constructed a method in which one maps measured outcome probabilities (primary data), which does not perfectly satisfy a strict equivalence constraint, into another set of probabilities (secondary data) that satisfies that equivalence constraint. Then, one uses the secondary data to calculate the parameter of interest in the experiment. We will use this method to strictly enforce the obliviousness constraint, and then compute $A_3$.

The primary data in our experiment consists of six $2\times 3$ matrices (one for each preparation $(x_0,x)$) with elements $\mathbf{P}^{x_0x}_{i,j}\equiv P^{Lab}(j|x_0,x,i)$ corresponding to performing measurement $i$ in the laboratory and obtaining outcome $j$. We will assume that the underlying physical theory governing the system is linear, allowing us to search for secondary data in the form of six other matrices $\{\mathbf{P'}^{x_0x}\}_{x_0,x}$ that are in the convex hull of $\{\mathbf{P}^{x_0x}\}_{x_0,x}$. That is; we let $\forall x_0,x: \mathbf{P'}^{x_0,x}=\sum_{x_0'=0}^2\sum_{x'=0}^{1} w^{x_0x}_{x_0',x'}\mathbf{P}^{x_0',x'}$, where $\forall x_0,x: w^{x_0x}_{x_0',x'}$ is a probability distribution. We seek secondary data which: (i) satisfies the obliviousness constraint, and (ii) on average is as close to the primariary data as possible. This corresponds to a linear program:
\begin{gather}\nonumber
S\equiv \max_{\{w\}} \frac{1}{6}\sum_{x_0=0}^2\sum_{x=0}^1 w^{x_0,x}_{x_0,x},\\
 \text{such that: }\sum_{x_0}\mathbf{P'}^{x_00}= \sum_{x_0}\mathbf{P'}^{x_01}.
\end{gather}
We find $S\approx 0.9938$, indicating that the secondary data is close to the primary data. Using the secondary data to compute $A_3$ we obtain $A_3^{sec} \approx 0.7118\pm 0.0365$. This is only marginally smaller than $A_3^{pri}$. It is in good agreement with the theoretical prediction of quantum theory and strictly satisfies the obliviousness constraint.

\textit{Conclusions.---} We have established relations between opertional statistics in a class of communication games and tests preparation contextuality. We showed close relations between quantum nonlocality and quantum correlations in such communication games, and also shown all mixed quantum states of finite dimension to be preparation contextual. Furthermore, we provided an experimental demonstration of a quantum communication game showing a large violation of a preparation noncontextuality inequality. 

We conclude with some open problems: 1) Do communication games without obliviousness constraints admit a connection to some operational physical assumption in the same spirit as presented here for games respecting an obliviousness constraint? 2) Are generalizations of the presented framework to more than two players possible? 3) Can the considered communication games be used in one-sided device independent cryptography protocols?

\textit{Acknowledgments.---}
The authors thank Ad\'an Cabello, Nicolas Gisin, Nicolas Brunner, Thiago Maciel and Artur Matoso for the useful discussions and comments. We extend particular gratitude to Debashis Saha and Anubhav Chaturvedi for enlightening comments and criticism. The project was financially supported by Knut and Alice Wallenberg foundation and the Swedish research council. A.T. acknowledges financial support from the Swiss National Science Foundation (Starting grant DIAQ). B.M is supported by FAPESP N$^{\circ}$ 2014/27223-2. 

\newpage

\title{Supplementary material: Communication games measure preparation contextuality}

\author{Alley Hameedi}\thanks{A. H. and A. T. contributed equally for this project.}
\affiliation{Department of Physics, Stockholm University, S-10691 Stockholm, Sweden}

\author{Armin Tavakoli}\thanks{A. H. and A. T. contributed equally for this project.}
\affiliation{Department of Physics, Stockholm University, S-10691 Stockholm, Sweden}
\affiliation{Groupe de Physique Appliqu\'ee, Universit\'e de Gen\`eve, CH-1211 Gen\`eve, Switzerland}

\author{Breno Marques}
\email{bmgt@if.usp.br}
\affiliation{Instituto de F\'isica, Universidade de S\~ao Paulo, 05315-970 S\~ao Paulo, Brazil}

\author{Mohamed Bourennane}
\affiliation{Department of Physics, Stockholm University, S-10691 Stockholm, Sweden}

\maketitle

\section{Preparation noncontextuality inequalities from parity-oblivious random access codes}

In this section, we provide an example of how our construction linking communication games to tests of preparation contextuality can be used. Specifically, we provide a family of preparation noncontextuality inequalities based on a broad family of communication games which can be understood as random access codes \cite{T15} supplemented with obliviousness constraints. The preparation noncontextuality inequalities of Ref.\cite{SB09} emerge as special instances of what follows.

Alice holds $x=x_1\ldots x_{n}\in \{0,\ldots,d-1\}^{\otimes n}=I_A$ with $p_A(x)=1/d^n$, and Bob holds $y\in\{1,\ldots,n\}$ with $p_B(y)=1/n$. For every $r=r_1\ldots r_n\in\{0,\ldots,d-1\}^n$ let the function $Z(r)=\sum_{i=1}^{n}\delta_{r_i,0}$ count the number of zero-elements in $r$, and consider the following sets $S_k^r=\{x|r\cdot x\equiv \sum_{i=1}^{n}x_ir_i=k\mod{d}\}$ for $k=0,\ldots, d-1$ and all strings $r$ with $Z(r)\leq d-2$. Impose an obliviousness constraint over the collection of these sets:
\begin{equation}\label{exob}
\forall k,r,k',r':	\frac{1}{q_{r,k}}\sum_{x\in S_k^r} p(b|x,y)=\frac{1}{q_{r',k'}}	\sum_{x\in S_{k'}^{r'}} p(b|x,y).
\end{equation}
For simplifying purposes, we will only consider values of $d$ that are prime numbers. This ensures that the equation $r\cdot x=k\mod{d}$ has exactly $d^{n-1}$ solutions for any relevant $r$ and $k$. Consequently, we have $q_{r,k}=q_{r',k'}$ and can thus ignore these factors in the obliviousness constraint \eqref{exob}. The obliviousness constraint is interpreted as Bob not being allowed to gain any information on the modulo $d$ sum (parity) of any weighted string of Alice's data that has length at least two. The specific game played by Alice and Bob is such that they earn a unit payoff if $b=x_y$ and otherwise no payoff. The preparation noncontextuality inequality for this family of parity-oblivious random access codes reads 
\begin{equation}\label{ex1}
A_{n,d}\equiv \frac{1}{nd^n}\sum_{y=1}^{n}\sum_{x\in \{0,\ldots,d-1\}^n} p(b=x_y|x,y)\leq \frac{n+d-1}{nd}.
\end{equation}
The left-hand-side is the performance in the communication game for given $(n,d)$, and the right-hand-side is the preparation noncontextual bound. We present the proof of this inequality at the end of this section. Notice however that a very simple preparation noncontextual strategy saturates the bound: let Alice always send $x_1$ to Bob. It is easily seen that the obliviousness constraint is satisfied. Whenever Bob has $y=1$, he outputs $b=x_1$ and earns a payoff, whereas if $y\neq 1$ he simply guesses the value of $x_y$, succeeding with probability $1/d$. Calculating this average returns the bound in Eq.\eqref{ex1}. It is interesting to note that the preparation noncontextuality inequalities derived in Refs.\cite{SB09} constitute special cases of the above corresponding to us setting $d=2$. 

For exploratory and examplifying purposes, we have performed numerical optimization $A_{n,d}$ with $(n,d)=(2,3)$ in a quantum model. The numerical optimization was implemented with a see-saw of semi-definite programs iterating between optimizations of the preparations of Alice and the measurements of Bob respectively, in each of which the obliviousness constraints acts as a convex constraints. These optimization procedures were carried out for a specific dimension $D$ of the Hilbert space in which Alice's preparations live. We have considered the cases of $D=3,\ldots,7$ and the results are presented in table \ref{tab:1}.
\begin{table}[h]
	\centering
	\begin{tabular}{| c| c| c| c| c| c|}
		\hline
		$ D: $ & $D=3$ & $D=4$ & $D=5$ & $D=6$ & $D=7$  \\ [1ex] 
		\hline 
		$A_{2,3}:$ & $0.6667$ & $ 0.6875$ & $0.6875$ & $0.6979$ & $0.6979$ \\ 
		\hline 
	\end{tabular}
	\caption{Numerical results for $A_{2,3}$ in a quantum model in which Alice communicates states of dimension $D$.}
	\label{tab:1}
\end{table}
We can numerical certify quantum preparation contextuality, i.e., a violation of the preparation noncontextuality inequality $A_{2,3}\leq 2/3$, for Alice communicating states of at least dimension $D=4$. An analogous numerical approach to $A_{3,3}$ with $D=4$ returned $A_{3,3}=0.5999$ violating the preparation noncontextual bound $5/9$. Our numerical are most likely not optimal, but they are sufficient to show the preparation contextuality of quantum theory can be detected by these inequalities. Also, our numerical results suggest that finding the optimal quantum violations of our preparation noncontextuality inequalities may be a non-trivial task which strongly depends on the dimension of Hilbert space. Exploring quantum violations of our preparation noncontextuality inequalities is left for future work.

\subsection{Proof of preparation noncontextuality inequalities \eqref{ex1}}
Here, we prove the preparation noncontextuality inequality \eqref{ex1}. As shown in the main text, the preparation noncontextual bound of $A_{n,d}$ is achieved by maximizing $A_{n,d}$ over all classical strategies of Alice and Bob that respect the obliviousness constraint, i.e., Alice encodes her data $x$ into a classical message $m$ which she sends to Bob who outputs $b$ by performing some computation on $m$ and $y$. The obliviousness constraint can be written
\begin{equation}\label{exobliv}
\forall k,r,k',r':	\sum_{x\lvert r\cdot x=k} p(m|x)=	\sum_{x\lvert r'\cdot x=k'} p(m|x).
\end{equation} 

Define the functions $\xi_r(x)=\omega^{r\cdot x}$ with $\omega=e^{2\pi i/d}$ mapping $\{0,\ldots,d-1\}^n$ onto $\{1,\ldots,\omega^{d-1}\}$. It is easily shown that 
$\sum_{x}\bar{\xi}_{r'}(x) \xi_r(x)=d^n\delta_{r,r'}$, where the bar denotes complex conjugation. Since $\{\xi_r\}_r$ form a complete orthogonal basis of our function space, we can perform a Fourier expansion of $p(m|x)$ as follows:
\begin{equation}\label{fourier}
p(m|x)=\sum_{r\in\{0,\ldots,d-1\}^n} c_r(m)\xi_r(x),
\end{equation}
where $c_r(m)$ are the Fourier coefficients which are obtained from 
\begin{equation}\label{fourier2}
c_r(m)=\frac{1}{d^n}\sum_{x\in\{0,\ldots,d-1\}^n}p(m|x)\bar{\xi}_{r}(x)=\frac{1}{d^n}\left[\sum_{x|r\cdot x=0}p(m|x)+\bar{\omega}\sum_{x|r\cdot x=1}p(m|x)+\ldots+\bar{\omega}^{d-1}\!\!\!\sum_{x|r\cdot x=d-1}p(m|x)\right].
\end{equation}
Enforcing the obliviousness constraint \eqref{exobliv}, and using that $1+\bar{\omega}+\ldots+\bar{\omega}^{d-1}=0$, we find that all coefficients $c_r(m)$ for which $Z(r)\leq d-2$ vanish. Only Fourier coeffients associated to strings $r$ that do not satisfy $Z(r)\leq d-2$ have non-zero values. There are $1+(d-1)n$ such strings: one is the all-zero string, and then there are strings such that the value $j\in\{1,\ldots,d-1\}$ appears at position $i\in\{1,\ldots,n\}$ while all other entries of $r$ are zero. For the all-zero string we name its Fourier coefficient $f^0(m)$, and for the string with entry $j$ at position $i$ with zeros everywhere else, we label the Fourier coeffient as $f^j_i(m)$. Hence, under the obliviousness constraint, Eq.\eqref{fourier} reduces to 
\begin{equation}\label{step}
p(m|x)=f^0(m)+\sum_{i=1}^{n}\sum_{j=1}^{d-1} f^j_i(m)\omega^{jx_i}.
\end{equation}
Let us write $\omega^{jx_i}=\sum_{l=0}^{d-1}\omega^{jl}\delta_{x_i,l}$ and insert this into \eqref{step} to obtain
\begin{equation}\label{fseries}
p(m|x)=f^0(m)+\sum_{i=1}^{n}\sum_{l=0}^{d-1} \delta_{x_i,l}\underbrace{\left[\sum_{j=1}^{d-1}\omega^{jl}f^j_i(m)\right]}_{\equiv t_{i}^{(l)}(m)}=f^0(m)+\sum_{i=1}^{n}\sum_{l=0}^{d-1} \delta_{x_i,l} t_{i}^{(l)}(m).
\end{equation}
For simplicity we will write $t_{i}^{(l)}(m)=t_i^{(l)}$, keeping in mind the dependence on $m$. An important property of the coefficients $t_i^{(l)}$ is the following:
\begin{equation}\label{sum0}
\forall i: \sum_{l=0}^{d-1}t_i^{(l)}=\sum_{j=1}^{d-1}\underbrace{\left[\sum_{l=0}^{d-1}\omega^{jl}\right]}_{=0 \text{ }\forall j} f^j_i(m)=0.
\end{equation}

We want to ensure that $ t_{i}^{(l)}$ are all real numbers. To that end, using \eqref{fourier2} we can evaluate $ t_{i}^{(l)}$ to
\begin{align}\nonumber
t_{i}^{(l)}=\frac{1}{d^n}\sum_{j=1}^{d-1}\sum_{k=0}^{d-1}\sum_{x|jx_i=k}\omega^{jl-k}p(m|x)=\frac{1}{d^n}&\sum_{j=1}^{d-1}\sum_{k=0}^{d-1}\sum_{x_i}\delta_{jx_i,k}\omega^{jl-k}\!\!\!\sum_{\substack{x_1\ldots x_{i-1}\\ x_{i+1}\ldots x_n}}p(m|x)\\
&=\frac{1}{d^n}\sum_{x_i}\left[\sum_{j=1}^{d-1}\omega^{j(l-x_i)}\right]\sum_{\substack{x_1\ldots x_{i-1}\\ x_{i+1}\ldots x_n}}p(m|x).
\end{align} 
Using geometric series, the sum in brackets can be shown to be equal to $d-1$ if $x_i=l$, and $-1$ otherwise. We obtain
\begin{equation}\label{step2}
t_{i}^{(l)}=\frac{d-1}{d^n}\sum_{x|x_i=l}p(m|x)-\frac{1}{d^n}\sum_{\substack{x\\ x_i\neq l}}p(m|x).
\end{equation}
Thus, $t_{i}^{(l)}$ are real numbers. A similar argument using \eqref{fourier2} shows that $f^0(m)=1/d^n\sum_{x}p(m|x)$ which is non-negative. 

We shall now map the expression in Eq.\eqref{fseries} into an expression of the form 
\begin{equation}\label{fseries2}
p(m\lvert x)=a_0(m)+\sum_{i=1}^{n}\left[a_{i,0}(m)\delta_{x_i,0}+\ldots +a_{i,d-1}(m)\delta_{x_i,d-1}\right],
\end{equation}
where all coefficients $a_0(m),a_{i,0}(m),\ldots,a_{i,d-1}(m)$ are non-negative real numbers. To achieve this mapping from \eqref{fseries} to \eqref{fseries2}, for each $i$ we break up Eq.\eqref{fseries} into different cases which we will treat separately. Each case is specified by two elements: (i)  the signs of the coefficients $\{t_i^{(l)}\}_l$, and (ii) the smallest number appearing in the set $\{t_i^{(l)}\}_l$. In general, up to relabelings, we can write each such case, for a specified value of $i$, as follows: $t_i^{(0)},\ldots,t_i^{(q-1)}\geq 0$ and  $t_i^{(q)},\ldots,t_i^{(d-1)}< 0$ for some $q$. We let the smallest element be $t_i^{(d-1)}$.  Notice that \eqref{sum0} forbids the cases in which all signs are positive or negative. 

We use \eqref{sum0} to write $t_i^{(s)}=t_i^{(s)}+\sum_{l=0}^{d-1}t_i^{(l)}$ for $s\neq d-1$. Thus,
\begin{equation}\label{decomp}
\sum_{l=0}^{d-1} \delta_{x_i,l} t_{i}^{(l)}= \sum_{l=0}^{d-2} \left(t_i^{(l)}+\sum_{k=0}^{d-2}t_i^{(k)}\right)\delta_{x_i,l}+t_i^{(d-1)}\underbrace{\sum_{l=0}^{d-1}\delta_{x_i,l}}_{=1}.
\end{equation}
Notice that
\begin{equation}
\forall l\in\{0,\ldots,d-2\}: t_i^{(l)}+\sum_{k=0}^{d-2}t_i^{(k)}\geq t^{(d-1)}_i+\sum_{k=0}^{d-2}t_i^{(k)}=0.
\end{equation}
Hence, the expression in the bracket in \eqref{decomp} is non-negative. For the given case, we define $a_{i,l}(m)=t_i^{(l)}+\sum_{k=0}^{d-2}t_i^{(k)}$ for $l=0,\ldots,d-2$ and $a_{i,d-1}=0$. Performing such a decomposition for each case, we can fully define our non-negative coefficients $a_{i,l}(m)$. In addition, each decomposition of the form \eqref{decomp} leaves us with a term that does not depend on $x$: in \eqref{decomp} it is $t_i^{(d-1)}$. When suming over all possible cases (and over $i$), we gather all these $x$-independent terms in one coefficient we call $a_0(m)$. We must show that $a_0(m)$ is non-negative. To achieve that, notice that for every case the $x$-dependent part of the corresponding equation \eqref{decomp} has at least one $l_i^*\in \{0,\ldots,d-1\}$ such that the coefficient in front of $\delta_{x_i,l_i^*}$ is zero. In Eq.\eqref{fseries} we have $l_i^*=d-1$ since $\delta_{x_i,d-1}$ does not appear. Define a string $\tilde{x}(m)$ that assigns the value $l_i^*$ to $\tilde{x}_i$, where $l_i^*$ is determined from the case associated to $m$. By construction, we find
\begin{equation}
\forall i: a_{i,0}(m)\delta_{x_i,0}+\ldots +a_{i,d-1}(m)\delta_{x_i,d-1}=0,
\end{equation}
and thus we have $p(m\lvert \tilde{x}(m))=a_0(m)$ which ensures that $a_0(m)$ is non-negative. This completes the transformation of \eqref{fseries} to \eqref{fseries2}.

Use normalization together with \eqref{fseries2} to obtain:
\begin{equation}
1=\sum_{m}p(m|x)=\sum_{m}a_0(m)+\sum_{i=1}^{n} \sum_{m} a_{i,x_i}(m).
\end{equation}
Define $G=\sum_{m}a_0(m)$ and $K_{i,x_i}\equiv \sum_{m} a_{i,x_i}(m)$. It follows that $K_{i,x_i}$ is independent of $x_i$ and thus $K_{i,0}=\ldots=K_{i,d-1}\equiv K_i$. Introduce re-labelings as follows
\begin{equation}
p(0)=G \hspace{7mm} p(i)=K_i \hspace{7mm} p_0(m)=\frac{a_0(m)}{p(0)} \hspace{7mm} p_{i,l}(m)=\frac{a_{i,l}(m)}{p(i)}.
\end{equation}
Here, $p(i)$ can be interpreted as a probability distribution over $i\in\{0,\ldots,n\}$, $p_0(m)$ is a probability distribution over $m$, and for each $i\in\{1,\ldots,n\}$ and $l\in\{0,\ldots,d-1\}$, $p_{i,l}(m)$ is a probability distribution over $m$. Hence, we may write \eqref{fseries2} as
\begin{equation}
p(m|x)=p(0)p_0(m)+\sum_{i=1}^{n}p(i)\sum_{l=0}^{d-1} p_{i,l}(m)\delta_{x_i,l}.
\end{equation}
This is understood as follows: Alice samples $i$ from $p(i)$, if she obtains $i=0$ she sends a message sampled from $p_0(m)$ whereas if she obtains $i\in\{1,\ldots,n\}$, she sends a message sampled from $p_{i,x_i}(m)$ where the choice of distribution is specified by her data element $x_i$. To find the optimal value of $A_{n,d}$ in \eqref{ex1}, note that Bob learns nothing about $x$ if $i=0$. Hence we put $p(0)=0$. In order for Bob to be able to distinguish the distributions $p_{i,0},\ldots,p_{i,d-1}$ from each other, their supports need to be disjoint, i.e., given any value of $m$ Bob can immedaitely determine from which distribution it was sampled. Therefore, when $i=y$, Bob always finds $b=x_i$. However, if $i\neq y$ Bob knows nothing about $x_i$ and thus only finds $b=x_i$ with probability $1/d$. The average value of $A_{n,d}$ is then
\begin{equation}
A_{n,d}=\frac{1}{n}+\frac{n-1}{n}\times \frac{1}{d}=\frac{d+n-1}{dn},
\end{equation}
which is the right-hand-side of Eq.\eqref{ex1}.

\section{Quantum strategy in communication game based on the CGLMP Bell inequality}
In the main text, we presented a preparation noncontextuality inequality based on the CGLMP Bell inequality for $d=3$ outcomes. That inequality was
\begin{equation}\label{app1}
A_3\equiv \frac{1}{12}\sum_{x_0,x,y,k}(-1)^k p(b=T_k|x_0,x,y) \leq 1/2,
\end{equation}
where $x=x_0x\in\{0,1,2\}\times \{0,1\}$ with $p(x_0x)=p(x_0|x)/2$ are associated to preparations of Alice, $y\in\{0,1\}$ with $p_B(y)=1/2$ is associated to measurements of Bob, and $T_k=x_0-(-1)^{x+y+k}k-xy\mod{3}$ for $k=0,1$. In quantum theory, the preparations $\rho_{x_0x}$ must be such that the following obliviousness constraint is satisfied;
\begin{equation}
\sum_{x_0=0}^{2}p(x_0|x=0)\rho_{x_00}=\sum_{x_0=0}^{2}p(x_0|x=1)\rho_{x_01}.
\end{equation}

We consider an explicit quantum strategy in this game. This strategy will be based on the maximal quantum violation of the CGLMP Bell inequality. First, we summarize this well-known strategy \cite{AD02}, and then we use the method in the main text to map a quantum strategy in the Bell test into a quantum strategy for the associated communication game.

The CGLMP Bell inequality for $d=3$ outputs reads
\begin{multline}\label{cglmp}
I^{cglmp}\equiv \frac{1}{4}\Big(\left[P(A_0=B_0)+P(B_0=A_1+1)+P(A_1=B_1)+P(B_1=A_0)\right]\\
-\left[P(A_0=B_0-1)+P(B_0=A_1)+P(A_1=B_1-1)+P(B_1=A_0-1)\right]\Big)\leq 1/2,
\end{multline}
where expressions inside $P(\cdot)$ are evaluated modulo $3$.

In order to maximally violate the CGLMP Bell inequality we let Alice and Bob share an entangled state $|\phi\rangle=\frac{1}{\sqrt{N}}\sum_{k=0}^{2}\gamma_k |kk\rangle$ with $\gamma_0=\gamma_2=1$ and $\gamma_1=\left(\sqrt{11}-\sqrt{3}\right)/2$, and $N=2+\gamma_1^2$ being a normalization. The associated density matrix is $\rho^{AB}=\ket{\phi}\bra{\phi}$. Alice has two measurement options, indexed by $X\in\{0,1\}$, given by $|a\rangle_{{X},A}=\frac{1}{\sqrt{3}}\sum_{k=0}^{2}\omega^{k(a+\alpha_{X})}|k\rangle$ for $a=0,1,2$ with $\alpha_0=0$ and $\alpha_1=1/2$, where we have defined $\omega=e^{2\pi i/3}$. The associated projection operator is $A_{a}^{X}=|a\rangle_{{X},A}\langle a|_{X,A}$. Similarly, Bob has two measurement options $|b\rangle_{{Y},B}=\frac{1}{\sqrt{3}}\sum_{k=0}^{2}\omega^{k(-b+\beta_{Y})}|k\rangle$ for $b=0,1,2$, $Y\in\{0,1\}$ and $\beta_Y=(-1)^Y1/4$. The associated projection operator is written $B_b^Y=|b\rangle_{{Y},B}\langle b|_{{Y},B}$. This quantum strategy yields the maximal violation $I^{cglmp}=\left(3+\sqrt{33}\right)/12\approx 0.7287$ \cite{AD02}.

We transform this quantum strategy in the Bell test to a quantum strategy for the communication game associated to the CGLMP Bell inequality. First, it is straightforwardly calculated that in the Bell test the marginal probabilities of Alice are uniform;
\begin{equation}
\forall a,X: P(a|X)=\Tr\left(A_{a}^X\otimes \mathbf{1} \rho^{AB}\right)=\frac{1}{3}.
\end{equation}
Hence, the preparations of Alice in  the communication game are defined to be
\begin{equation}\label{app2}
\rho_{x_0x}=3\Tr_A\left(A_{x_0}^x\otimes \textbf{1}\rho^{AB}\right).
\end{equation}
They satisfy the obliviousness constraint since
\begin{equation}
\forall x: \frac{1}{3}\sum_{x_0=0}^{2}\rho_{x_0x}=\Tr_A\left(\sum_{x_0=0}^{2} A_{x_0}^x\otimes \textbf{1}\rho^{AB}\right)=\Tr_A(\rho^{AB})=\rho^{B},
\end{equation}
which is independent of Alice measurement $x$.

The explicit form of Alice's preparations is obtained from expanding Eq.\eqref{app2} to 
\begin{equation}
\rho_{x_0x_1}=\frac{1}{N}\sum_{j,k=0}^{2}\gamma_j\gamma_k\omega^{(k-j)(x_0-\delta_{x,1}+\alpha_{x})}|k\rangle\langle j|,
\end{equation}
where we have additionally let $x_0\rightarrow x_0-1$ whenever $x=1$. That is no more than relabeling Alice's preparations. Letting Bob perform the measurements used to maximally violate the CGLMP Bell inequality, we find that the probability distribution of Bob's outcome is 
\begin{equation}
p(b|x_0x,y)=\frac{1}{3N}\sum_{k,j=0}^{2} \gamma_k\gamma_j\omega^{(k-j)(x_0-b+\alpha_{x}+\beta_y-\delta_{x,1})}.
\end{equation}
For the probabilities of our interest, as specified in Eq.\eqref{app1}, put $b=T_q$ for $q=0,1$. The resulting distribution does not depend on $x_0$. The final quantum performance can be written 
\begin{equation}\label{effcglmp}
A_3=
\frac{1}{3N}\sum_{k,j=0}^{2}\gamma_k\gamma_j \Bigg[\cos\left(\frac{\pi}{6}(k-j)\right)-\cos\left(\frac{\pi}{2}(k-j)\right)\Bigg]=\frac{3+\sqrt{33}}{12}.
\end{equation}
As expected, we have found that the quantum strategy maximally violating the Bell inequality \eqref{cglmp} maps into a quantum strategy giving an equivalent violation of our preparation noncontextuality inequality \eqref{app1}.

\section{Experimental Settings}
Table I presents the required parameters to prepare the desired input states $\psi_{jk} (j \in \lbrace1,2\rbrace, k \in \lbrace1,2,3\rbrace)$. Tables II-IV present the experimentally estimated probabilities corresponding to the detection events in the detectors $|\alpha\rangle$, $|\beta\rangle$ and $|\gamma\rangle$. 
\begin{table}[h]
	\centering
	
	\begin{tabular}{|c|c|c|c|c|c|c|}\hline
		\centering
		State & $\chi_1$ & $\chi_2$ & \multicolumn{2}{|c|}{Basis 1} & \multicolumn{2}{|c|}{Basis 2} \\ \hline\cline{1-7}
		\: $\psi_{11}$ & 77.01 & 24.93  & $\theta_1$  & 0  & $\theta_1$  & 22.5\\\hline
		\: $\psi_{12}$ & 12.98 & 20.07  & $\theta_2$  & 0  & $\theta_2$  & 9.73\\\hline
		\: $\psi_{13}$ & 36.80 & 34.79  & $\theta_3$  & 0  & $\theta_3$  & 22.5\\\hline
		\: $\psi_{21}$ & 54.78 & 81.28  & $\phi_1$  & 0  & $\phi_1$  & 45\\\hline
		\: $\psi_{22}$ & 53.19 & 10.21  & $\phi_2$  & 0  & $\phi_2$  & 45\\\hline
		\: $\psi_{23}$ & 54.78 & 53.71  & $\phi_3$  & 0  & $\phi_3$  & 0\\\hline

	\end{tabular}
	\caption{The orientation of $\chi_1$ and $\chi_2$ allow us to prepare any of the states $\psi_{jk} (j \in \lbrace1,2\rbrace, k \in \lbrace1,2,3\rbrace)$. Measurements in basis 1 (computational basis) and basis 2 are performed with the given $\theta_i$ and $\phi_i$ settings  ($i\in \lbrace1,2,3\rbrace$).}
	\label{tab1}
\end{table}

\begin{table}[h]
	\centering
	\resizebox{\textwidth}{!}{%
		\begin{tabular}{|c|c|c|c|c|c|c|}\hline
			\centering
			State & Basis 1-Proj.1 ($|\alpha\rangle$) & Basis 1-Proj.2 ($|\beta\rangle$) & Basis 1-Proj.3 ($|\gamma\rangle$) & Basis 2-Proj.1 ($|\alpha\rangle$) & Basis 2-Proj.2 ($|\beta\rangle$) & Basis 2-Proj.3 ($|\gamma\rangle$) \\  \hline\cline{1-7} 
			\: $\psi_{11}$ & 0.8191 $\pm$	0.0028 &	0.0775 $\pm$ 0.0011&	0.1034 $\pm$	0.0016&		0.7849 $\pm$	0.0025&	0.1521 $\pm$	0.0011&	0.063 $\pm$	0.0015
			\\\hline
			\: $\psi_{12}$ & 0.1402 $\pm$	0.0025 & 0.7826	$\pm$ 0.0022 &	0.0772 $\pm$	0.0012	&	0.0884 $\pm$	0.0014&	0.7672 $\pm$	0.0018	&0.1443 $\pm$	0.0026
			\\\hline
			\: $\psi_{13}$ & 0.069	$\pm$ 0.0015&	0.1234 $\pm$	0.0037&	0.8075 $\pm$	0.0019&		0.1477 $\pm$	0.0027& 0.0484 $\pm$	0.0017 &	0.8039 $\pm$	0.0032
			\\\hline
			\: $\psi_{21}$ & 0.8071	$\pm$0.0018 & 0.0992 $\pm$	0.0036&		0.0937 $\pm$	0.0016	&	 0.0856 $\pm$	0.0019& 	0.1035 $\pm$	0.0024&	0.8109 $\pm$	0.0033
			\\\hline
			\: $\psi_{22}$ & 0.0959	$\pm$ 0.0013&	0.7969 $\pm$	0.0022&	0.1072 $\pm$	0.0036&		0.8017 $\pm$	0.0027&	0.0941 $\pm$	0.0017	&0.1042 $\pm$	0.0031
			\\\hline
			\: $\psi_{23}$ & 0.1258 $\pm$	0.0016 &	0.061	$\pm$0.0012&	0.8132 $\pm$	0.0039&  0.1306 $\pm$	0.0019 & 0.7675 $\pm$	0.0031 & 0.1019 $\pm$	0.0023 
			\\\hline

		\end{tabular}}
		\caption{The estimated probabilities from the detection events in the single photon detectors $|\alpha\rangle$, $|\beta\rangle$ and $|\gamma\rangle$ along with the corresponding uncertainties for the \textit{two protocol measurement bases}. The reported errors include the poissonian and systematic errors. }
		\label{tab2}
	\end{table}
	
	\begin{table}[h]
		\centering
		\resizebox{\textwidth}{!}{%
			\begin{tabular}{|c|c|c|c|c|c|c|}\hline
				\centering
				State & Basis 1-Proj.1 ($|\alpha\rangle$) & Basis 1-Proj.2 ($|\beta\rangle$) & Basis 1-Proj.3 ($|\gamma\rangle$) & Basis 2-Proj.1 ($|\alpha\rangle$) & Basis 2-Proj.2 ($|\beta\rangle$) & Basis 2-Proj.3 ($|\gamma\rangle$) \\  \hline\cline{1-7} 
				\: $\psi_{11}$ & 0.5055	$\pm$ 0.0032&	0.3335	$\pm$	0.0029 &	0.161	$\pm$	0.0020&		0.7154	$\pm$	0.0036&	0.238	$\pm$	0.0026&	0.0466	$\pm$	0.0018
				\\\hline
				\: $\psi_{12}$ & 0.1566	$\pm$	0.0017&	0.6306	$\pm$	0.0035&	0.2128	$\pm$	0.0059&		0.0374	$\pm$	0.0008&	0.5732	$\pm$	0.0023&	0.3893	$\pm$	0.0039
				\\\hline
				\: $\psi_{13}$ & 0.254	$\pm$	0.0017&	0.0768	$\pm$	0.0062&	0.6693	$\pm$	0.0051&		0.2989	$\pm$	0.0011&	0.2151	$\pm$	0.0048&	0.486	$\pm$	0.0036
				\\\hline
				\: $\psi_{21}$ & 0.1261	$\pm$ 0.0028&	0.2491	$\pm$	0.0016&	0.6248	$\pm$	0.0057&		0.0641	$\pm$	0.0041&	0.3642	$\pm$	0.0024&	0.5717	$\pm$	0.0033
				\\\hline
				\: $\psi_{22}$ & 0.4848	$\pm$	0.0016&	0.1438	$\pm$	0.0012&	0.3715	$\pm$	0.0035&		0.7234	$\pm$	0.0040&	0.0609	$\pm$	0.0026&	0.2157	$\pm$	0.0028
				\\\hline
				\: $\psi_{23}$ & 0.2492	$\pm$	0.0015&	0.6564	$\pm$	0.0035&	0.0944	$\pm$	0.0013&		0.2977	$\pm$	0.0042&	0.5061	$\pm$	0.0039&	0.1962	$\pm$	0.0027
				\\\hline
				
			\end{tabular}}
			\caption{The estimated probabilities from the detection events in the single photon detectors $|\alpha\rangle$, $|\beta\rangle$ and $|\gamma\rangle$ along with the corresponding uncertainties for the \textit{first two tomographic bases}. The reported errors include the poissonian and systematic errors. }
			\label{tab3}
		\end{table}
		
		\begin{table}[h]
			\centering
			\begin{tabular}{|c|c|c|c|}\hline
				\centering
				State & Basis 3-Proj.1 ($|\alpha\rangle$) & Basis 3-Proj.2 ($|\beta\rangle$) & Basis 3-Proj.3 ($|\gamma\rangle$) \\  \hline\cline{1-4} 
				\: $\psi_{11}$ & 0.2614	$\pm$	0.0024&	0.4107	$\pm$	0.0047&	0.3279	$\pm$	0.0034
				\\\hline
				\: $\psi_{12}$ & 0.234	$\pm$	0.0017&	0.4775	$\pm$	0.0030&	0.2885	$\pm$	0.0042
				\\\hline
				\: $\psi_{13}$ & 0.2428	$\pm$	0.0030&	0.2196	$\pm$	0.0035&	0.5376	$\pm$	0.0041
				\\\hline
				\: $\psi_{21}$ & 0.2119	$\pm$	0.0042&	0.2886	$\pm$	0.0027&	0.4995	$\pm$	0.0064
				\\\hline
				\: $\psi_{22}$ & 0.2637	$\pm$	0.0036&	0.303	$\pm$	0.0027&	0.4333	$\pm$	0.0035
				\\\hline
				\: $\psi_{23}$ & 0.2455	$\pm$	0.0034&	0.538	$\pm$	0.0029&	0.2165	$\pm$	0.0029
				\\\hline
				
			\end{tabular}
			\caption{The estimated probabilities from the detection events in the single photon detectors $|\alpha\rangle$, $|\beta\rangle$ and $|\gamma\rangle$ along with the corresponding uncertainties for the \textit{final tomographic basis}. The reported errors include the poissonian and systematic errors. }
			\label{tab4}
		\end{table}
		


\begin{thebibliography}{99}


\bibitem{IC09}
M. Paw\l{}owski, T. Paterek, D. Kaszlikowski, V. Scarani, A. Winter, and M. \.{Z}ukowski,
Nature (London) \textbf{461}, 1101 (2009).



\bibitem{GH14}
A. Grudka, K. Horodecki, M. Horodecki, W. K\l{}obus, and M. Paw\l{}owski,
Phys. Rev. Lett. \textbf{113}, 100401 (2014).

\bibitem{BB15}
M. Banik, S. S. Bhattacharya, A. Mukherjee, A. Roy, A. Ambainis, and A. Rai,
Phys. Rev. A \textbf{92}, 030103(R) (2015).

\bibitem{GB10}
R. Gallego, N. Brunner, C. Hadley, and A. Ac\'in. 
Phys. Rev. Lett. \textbf{105}, 230501 (2010).

\bibitem{HG12}
M. Hendrych, R. Gallego, M. Micuda, N. Brunner, A. Ac\'in, and J. P. Torres,
Nature Physics \textbf{8}, 588-591 (2012).

\bibitem{AB12}
J. Ahrens, P. Badziag, A. Cabello, and M. Bourennane,
Nature Physics \textbf{8}, 592-595 (2012).
	
	\bibitem{PB11}
	M. Paw\l{}owski, and N. Brunner,
	 Phys. Rev. A \textbf{84}, 010302(R) (2011).
	 
	\bibitem{LP12}
	H-W. Li, M. Paw\l{}owski, Z-Q. Yin, G-C. Guo, and Z-F. Han.
	Phys. Rev. A \textbf{85}, 052308 (2012).
	
	\bibitem{AB14}
	V. D'Ambrosio, F. Bisesto, F. Sciarrino, J. F. Barra, G. Lima, and A. Cabello,
	Phys. Rev. Lett. \textbf{112}, 140503 (2014).
		
	
	\bibitem{T15}
	A. Tavakoli, A. Hameedi, B. Marques, and M. Bourennane,
	Phys. Rev. Lett. \textbf{114}, 170502 (2015).
	


	\bibitem{SB09}
	R. W. Spekkens, D. H. Buzacott, A. J. Keehn, B. Toner, and G. J. Pryde,
	Phys. Rev. Lett. \textbf{102}, 010401 (2009).	

	\bibitem{CK16}
	A. Chailloux, I. Kerenidis, S. Kundu, and J. Sikora,
	New J. Phys. \textbf{18}, 045003 (2016).

	\bibitem{A16}
	A. Ambainis, M. Banik, A. Chaturvedi, D. Kravchenko, and A. Rai,
	arXiv:1607.05490.		
	
	\bibitem{Sp05}
	R. W. Spekkens,
	Phys. Rev. A \textbf{71}, 052108 (2005).
	
	
\bibitem{S08}	
Robert W. Spekkens,
Phys. Rev. Lett. \textbf{101}, 020401 (2008).


\bibitem{P14}
M. F. Pusey,
Phys. Rev. Lett. \textbf{113}, 200401 (2014).

	\bibitem{LM13}
	M. S. Leifer, and O. J. E. Maroney,
	Phys. Rev. Lett. \textbf{110}, 120401 (2013).


	\bibitem{KS15}
	R. Kunjwal, and Robert W. Spekkens,
	Phys. Rev. Lett. \textbf{115}, 110403 (2015).		



\bibitem{Supplementary}
See Supplementary Material.


\bibitem{CHSH69}
J. F. Clauser, M. A. Horne, A. Shimony, and R. A. Holt,
Phys. Rev. Lett. \textbf{23}, 880 (1969).


	\bibitem{TM16}
	A. Tavakoli, B. Marques, M. Paw\l{}owski, and M. Bourennane,
	Phys. Rev. A \textbf{93}, 032336 (2016).


	\bibitem{M14}
	M. Banik, S. S. Bhattacharya, S. K. Choudhary, A. Mukherjee, and A. Roy,
	Foundations of Physics \textbf{44}, 1230-1244 (2014).


	\bibitem{CGLMP02}
	D. Collins, N. Gisin, N. Linden, S. Massar, and S. Popescu,
	Phys. Rev. Lett. \textbf{88}, 040404 (2002).	


	

	\bibitem{chen}
	J-L. Chen, D-L. Deng, and M-G. Hu,
	Phys. Rev. A \textbf{77}, 060306(R) (2008).
	
	
	\bibitem{AD02}
	A. Acin, T. Durt, N. Gisin, J. I. Latorre,
	Phys. Rev. A \textbf{65}, 052325 (2002).


		

\bibitem{hw}
B. G. Englert, C. Kurtsiefer, and H. Weinfurter,
Phys. Rev. A \textbf{63}, 032303 (2001).	
	
\bibitem{VQT1}
T. O. Maciel, A. T. Ces\'{a}rio, and R. O. Vianna,
Int. J. Mod. Phys. C, \textbf{22}, 1361 (2011).

\bibitem{VQT2}
D. S. Gon\c{c}alves, C. Lavor, M. A. Gomes-Ruggiero, A. T. Ces\'ario, R. O. Vianna, and T. O. Maciel,
Phys. Rev. A \textbf{87}, 052140 (2013)

	\bibitem{MP16}
	M. D. Mazurek, M. F. Pusey, R. Kunjwal, K. J. Resch, and R. W. Spekkens, 
	Nat. Commun. \textbf{7}, 11780 (2016).
	

\end{thebibliography}

\begin{thebibliography}{99}
		
		
		\bibitem{T15}
		A. Tavakoli, A. Hameedi, B. Marques, and M. Bourennane,
		Phys. Rev. Lett. \textbf{114}, 170502 (2015).
		
		
		\bibitem{A16}
		A. Ambainis, M. Banik, A. Chaturvedi, D. Kravchenko, and A. Rai,
		arXiv:1607.05490.	
		
		
		
		\bibitem{SB09}
		R. W. Spekkens, D. H. Buzacott, A. J. Keehn, B. Toner, and G. J. Pryde,
		Phys. Rev. Lett. \textbf{102}, 010401 (2009).
		
		\bibitem{AD02}
		A. Acin, T. Durt, N. Gisin, J. I. Latorre,
		Phys. Rev. A \textbf{65}, 052325 (2002).
		
		
		
		\end{thebibliography}
\end{document}